\documentstyle[12pt,fleqn,epsf,epsfig]{article}

\begin{document}
\title{Sympatric speciation: compliance with phenotype diversification from a single
genotype}
\author{
        Kunihiko Kaneko\\
        {\small \sl Department of Pure and Applied Sciences}\\
        {\small \sl University of Tokyo, Komaba, Meguro-ku, Tokyo 153, JAPAN}\\
       and\\
        Tetsuya Yomo\\
        {\small \sl Department of Biotechnology, Faculty of Engineering}\\
        {\small \sl Osaka University 2-1 Suita, Osaka 565, JAPAN}
\\}
\date{}
\maketitle

\vspace{.1in}

\hspace{.3in} Key words:  sympatric speciation, differentiation, symbiosis, isologous diversification

\vspace{.1in} 
\begin{abstract}
A novel mechanism for sympatric speciation that
takes into account complex bio-processes within
each individual organism is proposed.
According to dynamical systems theory, organisms with identical
genotypes can possess differentiated physiological states and may
coexist  `symbiotically' through appropriate mutual interaction. 
With mutations, the phenotypically differentiated
organisms gradually come to possess distinct genotypes, while maintaining 
their symbiotic relationship.  This
symbiotic speciation is robust against sexual recombination, because
offspring of mixed parentage, with intermediate genotypes, are less fit
than their parents. This leads to sterility of the hybrid.
Accordingly, a basis for mating preference also arises.
\end{abstract}

\pagebreak

The question posed by Darwin(1859) of why organisms are separated into distinct
groups, rather than exhibiting a continuous range of characteristics,  has 
not yet been fully answered.  In spite of several explanations involving
sympatric speciation,
``we are not aware of any explicit model demonstrating the instability of a
sexual continuum," according to Maynard-Smith and Szathmary(1995).
The difficulty involving stable sympatric speciation is that it
is not clear how two groups,
which have just begun to separate, coexist while mutually interacting.
Here, we propose a mechanism through which two groups with
little (or no) difference in genotype form  a `symbiotic' relationship under
competition.
This mechanism is understood in terms of the `isologous diversification
theory'(Kaneko \& Yomo 1997,1999),
according to which organisms with identical genotypes spontaneously split
into distinct phenotypes and establish a relationship 
in which the existence of one group is mutually supported by the other.
By considering genetic mutations and sexual recombinations,
a sympatric speciation process follows, resulting in the formation
of distinct genotypic groups exhibiting reproductive isolation. 
This process is robust with respect to fluctuations, due to the symbiotic
relationship.  The hybrid offspring of
the two groups becomes sterile, and also provides a
basis for mating preference, a major mechanism in sympatric
speciation(Maynard-Smith 1966, Lande 1981, Turner \& Burrows 1995, Kondrashov \& Kondrashov,
1999, Dieckmann  \& Doebeli 1999, Futsuyma 1986, Howard \& Berlocher ed. 1998).

To study phenotypic and genotypic diversification through
interaction, we have to consider a developmental process that maps a
genotype to a phenotype.  As an illustrative model, we consider a dynamic
process consisting of several interacting metabolic cycles.
Each organism possesses such internal dynamics with several metabolic
cycles, while it selectively consumes external resources, 
depending on their internal dynamics, and transforms them into some products.  
Through this process, organisms mature and eventually become ready for reproduction.

For most studies on population biology and evolution so far, it is widely assumed that
a phenotype is uniquely determined for a given genotype and environment.
If this assumption were always true, population dynamics only of genotypes
would be sufficient to study the evolutionary process theoretically.  
However, there are cases that organisms of the same genotype
may take distinct phenotypes through interaction.

First, some mutant genotypes related to malfunctions show various phenotypes, 
each of which appears at a low probability(Holmes 1979). This phenomenon is 
known as low or incomplete penetrance(Opitz 1981), which suggests plastic ontogenesis. 

Although the origin of low penetrance in multicellular organisms
may not be well clarified, differentiation of physiological
states is already known in bacteria (Novick and Weiner, 1957).
Furthermore, one of the authors and his colleagues have
found that specific mutants of {\it E. coli} 
show (at least) two distinct types of enzyme activity, although they 
have identical genes.  These different types 
coexist in an unstructured environment of a chemostat (Ko et al. 1994), and
this coexistence is not due to spatial localization.  Coexistence of 
each type is supported by each other.  Indeed, when one type of {\it E. coli}
is removed externally, the remained type starts differentiation again
to recover the coexistence of the two types.
In addition, even at a molecular level, a
mutant gene of xylanase was shown to produce various levels of enzyme
activity(Ko et al., 1994). A mechanism for a single gene to show various 
levels of molecular function has also been elucidated in physicochemical terms
(Kobayashi et al. 1997).

Such differentiation of phenotype is also discussed as
a possibility of different inheritable states of a same genotype 
(see e.g., Landman 1991).
Although we do {\bf not} assume any epigenetic inheritance here since
its relevance to evolution is still controversial, it should be noticed 
that the existence of different physiological states from a single genotype 
itself is demonstrated experimentally, even if one does not accept the
inheritance of the states.

Note also that the differentiation of phenotypes from the same genotype 
is supported also theoretically, as will be mentioned later.
Here we will study the relevance 
of such phenotypic diversification to evolution.

{\bf Model}

In our theoretical model, the phenotype is represented by a set of
variables, corresponding to metabolic processes or some other biological
processes.  
To be specific, each individual $i$
has several (metabolic or other) cyclic processes, and the
state of the $j$-th process 
at time $n$ is given by $X^j_n (i)$.  With $k$ such processes, 
the state of an individual is given by the set
$(X^1_n (i)$,$X^2_n(i)$, $\cdots ,X^k_n(i))$, which defines the phenotype.
This set of variables can be regarded as concentrations of
chemicals, rates of metabolic processes, or some
quantity corresponding to a higher function.
The state changes temporally according to
a set of deterministic equations with some parameters. 

Since genes are simply chemicals contained in DNA, they
could in principle be included in the set of variables.  However,
according to the `central dogma of molecular biology'(Alberts et al. 1994), the gene has a
special role among such variables:  Genes can affect phenotypes, the set
of variables, but the phenotypes cannot directly change the code of genes. 
During one generation, changes in genes are negligible compared with those of
the phenotypic variables they control. Hence, in our model, the set
corresponding to genes is represented by parameters that govern the
dynamics of phenotypes, since the parameters in an equation are not
changed while they control the dynamics of the variables.

Our model consists of the following procedures.

(i) Dynamics of the phenotypic state:
The dynamics of the variables $X_n^j(i)$ consist of a mutual influence
of cyclic processes ($X_n^{\ell}(i)$) and interaction with other organisms 
($X_n^j(i')$).

 (ii)  Growth and Death:
Each individual splits into two when a given condition for growth is satisfied.
Taking into account that the cyclic process corresponds to a metabolic,
genetic or other process that is required for reproduction, we assume
that the unit replicates when the accumulated number of cyclic processes goes
beyond some threshold.  To introduce competition for survival,
death is included both by random removal of organisms at some rate 
as well as by a given death condition based on their state. 

(iii) Genetic parameter and mutation:
Following the above argument, genes are represented as parameters
in the model. With reproduction, the genes  slowly mutate.  The set of parameters in the model changes slightly 
through mutation when offspring is reproduced.

To be specific we consider the following model.
First, the state variable $X_n^j(i)$ is split
into its integer part $R_n^{\ell}(i)$ and the
fractional part $x_{n}^{\ell}(i)=mod[X_{n}^{\ell}(i)]$.
The integer part $R_n^j(i)$ is assumed to give the number of times
the cyclic process has occurred since the individual's birth, while the
fractional part $x_{n}^{\ell}(i)$ gives the phase of oscillation in the
process.  
As a simple example, the internal dynamics of the cyclic process is assumed to be given by
$\sum_{m} \frac{a^{\ell m}}{2}sin(2\pi x_n^m(i))$, while the interaction
among organisms is
given by the competition for resources among the $N_n$ organisms existing
at the moment, 
given by $I^{\ell}(i)=p sin(2\pi x_n^{\ell}(i))+\frac{s^{\ell} 
-\sum_j p sin2\pi (x_n^{\ell}(j))}{N_n}$.
(The second term comes from the constraint $\sum_i
I^{\ell}(i)=s^{\ell}$, i.e., the condition that $N$ individuals compete
for a given resource $s^{\ell}$ at each time step.  The first term
represents the ability to secure the resource, depending on the state.)
Our model is given by

\begin{math}
X_{n+1}^{\ell}(i)=X_n^{\ell}(i)+ \sum_m \frac{a^{\ell m}(i)}{2} sin( 2\pi
x_n^m(i))-\sum_m \frac{a^{m \ell}(i)}{2} sin( 2\pi x_n^{\ell}(i)) \end{math}
\begin{equation}
+p sin(2\pi x_n^{\ell} (i))+\frac{s^{\ell} -\sum_j p sin2\pi
(x_n^{\ell}(j))}{N_n}. \end{equation}

Then, as a specific example, the condition for the reproduction is given by
\begin{math}
\sum_{\ell}R_n^{\ell}(i) \geq Thr.
\end{math}
The rotation number $R_n^{\ell}(i)$ is reset to zero when the corresponding 
individual splits.  On the other hand,
as the death condition,
an individual with $R_n^{\ell}(i) < -10$ (i.e., with a reverse process)
is removed.

Next, genotypes
are given by a set of parameters $a^{m \ell}(i)$, representing the 
relationship between the
two cyclic processes $\ell$ and $m$ ($1 \leq \ell,m \leq k$).  
With each division, the parameters $a^{m \ell}$ are changed to
$a^{m \ell}+\delta$ with $\delta$, a random number over $[-\epsilon,\epsilon]$,
with small $\epsilon$, corresponding to the mutation rate.  
Although the results in the figure adopt the mutation rate
$\epsilon =.001$, change of mutation rate is just responsible for the
speed of the separation of the parameters, and the conclusions are independent of
its specific value.

Let us make some remarks on our model.  Each term $a^{\ell m}sin (x_n^m(i))$ gives how  a process $x^m$ influences on $x^{\ell}$.  For example,
in a metabolic process, one cycle influences some other through catalytic reactions, 
depending on the activity of an enzyme corresponding to it.
With the change of genes, the activities of enzymes can change, which
leads to the change of the parameter values of $a^{\ell m}$ accordingly.  Following this argument, genotypes are regarded to be represented by a set of parameters $a^{\ell m}$.
Indeed, we have also studied a specific biochemical reaction network 
model with its catalytic efficiency as a genetic parameter, and the results to be presented 
are observed.

The interaction term $p sin(2\pi x_n^{\ell} (i))$ represents the influence on the cyclic process between individuals through 
the exchange of chemicals (or by other means).  Since this term can change 
its sign, chemicals can be secreted from each individual to the environment.  Then, the resources that
are taken out from one individual may be used by some other.  
Through this ecological interaction, individuals may keep some
relationship if they are differentiated.

Of course, the above explanation is just one example
of correspondence of our model to a real biological process.
As long as its mathematic structure
is common, the validity of our model is not restricted to the
above example, and the results to be presented can generally be applied.

{\bf Scenario for Symbiotic Sympatric Speciation}

The above model is one of the simplest to discuss 
loose developmental process.   We have also carried out simulations 
of several models of this type, for example, consisting of a metabolic
process of autocatalytic networks. 
Through the simulations and theoretical considerations of several models
the following mechanism for speciation is proposed.
Since the characteristic
features for speciation, to be presented as follows, are common,
we adopt the above simplest model to illustrate the scenario here.  Note, of course,
the scenario for the speciation is expected to work in a more
realistic model including much complicated processes for development
and interaction.

{\bf Stage-1: Interaction-induced phenotypic differentiation}

When there are many individuals interacting for finite resources, the
phenotypic dynamics begin to differentiate even though the
genotypes are identical or differ only slightly (see the `light blue points' 
in Fig.1. This
differentiation generally appears if nonlinearity is involved in the
internal dynamics of some phenotypic variables(Kaneko \& Yomo 1997,1999, Furusawa \& Kaneko 1998). Then, slight
phenotypic differences between individuals are amplified by the internal
dynamics (e.g., metabolic reaction dynamics). Through interaction
between organisms, the different phenotypic dynamics tend to be
grouped into two (or more) types, despite the fact that all
have identical (or only slightly different) genotypes.  
In the example of Fig.1a, the phenotype
splits into two groups, which we refer to as `upper' and `lower'
groups.

This differentiation process has recently been clarified as isologous
diversification(Kaneko \& Yomo 1997,1999), in which two groups with distinct
phenotypes appear even in a group with a single genotype.  This
interaction-induced phenotype diversification is a general consequence
when the developmental process with interaction between organisms is
considered as a nonlinear dynamics process(Kaneko 1991,1994).  When there is 
an instability in 
the dynamics, the temporal evolution of individuals  in phenotype space begins 
to diverge. Then, through interactions, 
these dynamics are stabilized through the formation of distinct groups 
with differentiated states in the
pheno-space.  The existence of the two (or more) groups eliminates the
instability in the dynamic (metabolic) process that exists when one of 
the groups is isolated.
Hence, the existence of all groups is required for the survival of each.
For example, if a group of one type is
removed, then the phenotype of individuals of another type changes
in compensation for the missing type.

To put the above explanation in biological terms, consider a given 
group of organisms
faced with a new environment and not yet specialized for the processing of
certain specific resources.
Each organism has metabolic (or other ) processes with a biochemical
network.  As the number of organisms increases, they compete for
resources.  The interaction, for example, results from the use of some 
byproduct of one organism by others.  As this interaction becomes stronger, the 
phenotypes become diversified to allow for different uses of metabolic cycles, 
and they split into two (or several) groups.  Each group is specialized in some
metabolic cycles, and also in the processing of some
resources.  Here, the byproduct of metabolic
processes of one group is necessary for another group to allow for it to
specialize in some particular
metabolic cycle. 
Resources secreted out from one group can be used
as a resource for the other group, and vice versa. 
Hence, the two groups realize a differentiation of
roles and form a symbiotic relationship. Each group is
regarded as specialized in a different niche, which is provided
by another group.

As an extreme case, this differentiation can occur even 
when a single resource is supplied externally (i.e., $s^j =0$ for $2 \leq j \leq k=3$).
In this case, the temporal average of $p sin(x^2_n(i))$ is positive 
for one group and negative for the other, while that of
$p sin(x^3_n(i))$ has the opposite sign.
With this differentiation of phenotypes,
one group uses $x^2$ as a resource for growth provided by the other,
which uses $x^3$ as a resource.

It should be pointed out that the the progeny of a reproducing individual 
belonging to one group may belong to the other group at this stage, since
all groups still have almost identical gene sets.

{\bf Stage-2: Co-evolution of the two groups to amplify the difference of
genotypes}

Now we discuss the evolutionary process of genotypes.  After the phenotype is
differentiated into two groups, the genotype (parameter) of each group
begins to evolve in a different direction, as shown in Fig.1 and Fig.2.  This
evolution occurs, since for the upper (lower) group, those individuals 
with a smaller (larger) parameter value reproduce faster.  In our numerical
simulations, there always exists (at least one) such parameter.  As a
simple illustration, assume that the two groups use certain metabolic 
processes differently. If the upper group uses one metabolic
cycle more, then a mutational change of the relevant parameter to
enhance this cycle is favored for the upper group, while a change to
reduce it (and enhance some others) may be favored for the lower group. 
In other words, each organism begins to adapt in one of the niches
formed by another (or others).

Note that $R^2$ also takes a different value between the two groups
in the other way round, since $\sum_j R^j=Thr$ for each division.
As for the parameter change, the values of 
$a^{12}$ and $a^{21}$ split first in this example, but then
$a^{23}$ and later other parameter values also start to split
into the two groups.  Several genes start to be responsible for the
differentiation.

As the genetic separation progresses, phenotypic differences
between the two groups also become amplified (see Fig.1).  With
the increase in the split in genotypes, it begins to become the case
that offspring of members of a given group
certainly keep the phenotype of this group.
Since the phenotype of one group stabilizes the other, the
evolutions of the two groups
are interdependent.  Hence the tempo of the genetic
evolution in one group is related to that of the other.  
The two groups co-evolve, maintaining their `symbiotic' relationship,
established in the previous stage.  Indeed, as shown in Fig.2,
the growth speeds of the two groups remain  of the same order, even if each
genotype and phenotype change with time.  

With the above described co-evolutionary process, the phenotype 
differentiation is
fixed to the genotype.  In much later generations, this fixation is
complete. In this case, even if one group is isolated, offspring with the
phenotype of the other group are no longer produced. 
Offspring of each group keep their phenotype (and genotype) on their own.

{\bf Reproductive Isolation with respect to Sexual Recombination}

The importance of the present scenario lies in the robustness of the  speciation
process.  Even if one group happens to
disappear through some fluctuations at the initial stage of the speciation process, 
coexistence of the two distinct phenotypic groups is recovered.
Hence, the present process is also expected to be
stable against sexual recombination, which mixes the two genotypes  
and may bring about a hybrid between the two genotypes.  
To demonstrate this stability, we have extended the previous model
to include this mixing of genotypes by sexual recombination.

Here, we have modified our model so that the sexual recombination occurs
to mix genes.  To be specific, the reproduction occurs when two
individuals $i_1$ and $i_2$ satisfy the threshold
condition ($\sum_{\ell}R^{\ell}_n(i_k)>Thr$), and then the two genotypes are 
mixed.  As an example we have produced two offspring $j=j_1$ and $j_2$,
from the individuals $i_1$ and $i_2$ as 

\begin{equation}
a^{\ell m}(j)=a^{\ell m}(i_1)r^{\ell m}_j+ a^{\ell m}(i_2)(1-r^{\ell m}_j) +\delta
\end{equation}
with a random number $0<r^{\ell m}_j<1$ 
to mix the parents' genotypes, besides 
the random mutation term by $\delta$.  
Even if two separated groups
may start to be formed according to our scenario,  the above recombination
forms `hybrid' offspring with intermediate
parameter values $a^{\ell m}$ when
two organisms from different groups mate.
Also, depending on the random number, for some offspring,
the parameter value $a^{lm}$ may be
closer to one of the parents, but that of $a^{l'm'}$ 
may be closer to that of the other of the parents.  Accordingly recombinations of the two group can 
lead to a different combination of alleles, since the two groups take different combination of the parameter values $\{a^{\ell m} \}$,
from the two groups.

Although the hybrid is formed in this random mating with some proportion
($1/2$ if the two groups have equal population),
it turns out that this hybrid, irrespective
of which phenotype it realizes, has a lower reproduction rate than the
other two groups which have a `matched' genotype-phenotype correspondence 
with a higher reproduction rate.
In Fig.3, we have plotted the average offspring number for given
genotype parameters (to be precise, the average over a given range of
parameters),  As shown, a drop at the intermediate value in the offspring 
number starts to appear, through generations.
Within few dozen generations, as certain genotypic
parameters are apart, there is little or no chance for a hybrid to
reproduce, and F1 sterility results.

Note that this conclusion is drawn 
even without assuming mating preference.  Rather, it is natural,
according to the present scenario, that 
mating preference in favor of similar phenotypes evolves, since it is 
disadvantageous for individuals to produce a sterile hybrid.
In other words,
the present mechanism also provides a basis for the evolution of sexual 
isolation through mating preference. Note, however, that in sympatric
speciation starting from only the mating preference, 
one of the groups may disappear due to fluctuations
when its population is not sufficiently large.  In contrast, according to our
scenario, the coexistence of the two groups is restored even under disturbances.  
Hence, it is concluded that our mechanism yields {\em robust}
sympatric speciation, i.e., differentiation of geno- and pheno-types and 
sexual reproductive isolation(Dobzhansky, 1951), even in the situation in which
all individuals interact with all others equally.

{\bf Importance of Phenotypic Differentiation}

Evolution according to our scenario often leads to specialization
with regard to resources through competition. Indeed, the
coexistence of two (or more) species
{\em after} the completion of the speciation is also supported by the resource 
competition theory of Tilman(1976,1981).  However, to realize the speciation 
{\em process}, phenotype differentiation from a single
genotype is essential.  As long as phenotype is uniquely determined by 
genotype, two individuals with a slight genotype difference can have only
a slight phenotype difference also.  Since competition is strong among 
individuals with similar phenotypes, they cannot coexist as a different group.
Hence two groups cannot be differentiated from a group of single (or similar) 
genotypes.  Contrastingly, in our scenario, even if the genotypes of two
individuals are the same or only slightly different, their phenotypes 
need not be similar, and can in fact be  
of quite different types, as shown in Fig.1.  Accordingly, these
two groups can coexist.

To check the importance of
this phenotypic differentiation from a single genotype, we
have also performed several numerical experiments with our model, by choosing
parameters so that differentiation into two distinct phenotype groups
does not occur initially.  In this case, separation into 
two (or more) groups with distinct pheno/geno-types is not observed,
even if the initial variance of
genotypes is large, or even if a large mutation rate is adopted.
This clearly demonstrates the relevance of phenotypic differentiation.

On the other hand, the genetic differentiation always occurs when the
phenotype differentiates into two (or more) groups. 
To be specific, in our model, the condition for the 
differentiation is as follows:
First, the parameter $p$ should be larger
than some value. For example, for $k=3$  with
$s^1=2$,$s^2=4$,$s^3=6$ and with the initial parameters $a^{\ell m}(i)
\approx -0.2/(2\pi)$, the differentiation appears for 
$p\stackrel{>}{\approx}1.8$.
Second, the resource term per unit ($\sum_j s^j/N$) should be smaller than
some threshold value. For example, the threshold resource is
$s_{thr}\approx 10$, for $s^1=s^2=s^3$, $p=1.5/(2\pi)$, $N \approx 300$ and
the initial parameters $a^{\ell m}(i)\approx -.1/(2\pi)$.
Note that these conditions imply strong interaction in competing for resources,
and are easier to be satisfied, 
as the number of individuals competing for given resources
increases.

{\bf Discussion}

In the symbiotic speciation process, the potentiality for a single genotype to
produce several phenotypes declines. After the
phenotypic diversification of a single genotype, each genotype again
appears through mutation and assumes one of the diversified phenotypes in the
population. Thus the one-to-many correspondence between the original
genotype and phenotypes eventually ceases to exist. As a result, one may see 
a single genotype expressing small numbers of phenotypes in nature, since most
organisms at the present time have gone through several speciation
processes.  One can also expect that mutant genotypes tend to have a
higher potentiality  than the wild-type genotype
to produce various phenotypes.
Indeed, this expectation is consistent with the observation that
low or incomplete penetrance(Opitz 1981, Holmes 1979) is
more frequently observed in mutants than in a wild type.

Taking our results and experimental facts into account, one can predict
that new species or organisms emerging as a species have a high
potentiality to produce a variety of phenotypes, while `living fossils',
such as {\it Latimeria chalumnae} and {\it Limulus}, have a
stable expression of a small number of phenotypes.
Relationship between evolvability and plasticity in
ontogenesis  is an important topic to be pursued.

Since the speciation discussed in this paper is triggered by interaction,
and not merely
by mutation, the process is not so much random as deterministic. In fact,
the speciation process occurs irrespectively of the adopted random number in the
simulation. Some of
the phenotypic explosions in nature that have been
recorded as occurring within short geologic periods, may have
followed the deterministic and relatively fast process of interaction-induced
speciation.  
Hence, our scenario may shed some light on the variation of 
timescales on which evolution  proceeds, e.g., 
punctuated equilibrium(Gould \& Eldredge 1977).
Here it should be noted that the change in phenotypes occur in few generations.
The speed of genetic change, of course, depends on the
mutation rate, but the present mechanism is found to work even for any
smaller mutation rate (say $\epsilon =10^{-6}$).

In the present paper, we have mostly reported the case only with 3 process
($k=3$), but we have numerically
confirmed that the present speciation process works also  for $k>3$
(e.g., $k=10$).
By choosing a model with many cyclic processes, we have also found successive
speciation of genotypes into several groups from a single genotype.  
With evolution, the phenotypes begin to be separated into two groups, 
each of which is specialized in some processes, and depends on the byproducts 
of the other.  Later, the 
species diverge into further specialized groups, which are
fixed into genotypes.  This process is relevant to consider
adaptive radiation.

Discussion of the mechanism involved in evolution often consists of mere
speculation.  Most important in our scenario, in contrast, is its
experimental verifiability. Isologous diversification has already been
observed in the differentiation of enzyme activity of {\it E. coli} with
identical genes(Ko et al. 1994).  By observing the evolution of {\it E. coli} in
the laboratory(Xu et al., 1996, Kashiwagi et al., 2000), controlling the strength of the 
interaction through the population density, one can check if the evolution on 
genetic level is accelerated through
interaction-induced phenotypic diversification. Our isologous symbiotic
speciation, based on dynamical systems theory, numerically confirmed
and biologically plausible, can be verified experimentally.

\vspace{.2in}

Acknowledgment:
We thank M. Shimada for
illuminating comments.  The present work is supported by 
Grants-in-Aids for Scientific Research from
the Ministry of Education, Science and Culture of Japan
(Komaba Complex Systems Life Project).


\vspace{.5in}

{\bf References}
\begin{itemize}

\item
B. Alberts et al.,
{\sl The Molecular Biology of the Cell}(Garland, New York ed.3. 1994).

\item
C.Darwin,
{\sl On the Origin of Species by means of natural selection or
the preservation of favored races in the struggle for life}
(Murray,London,1859).

\item
U.Dieckmann and M.Doebeli, On the origin of species by sympatric speciation,
{\sl Nature}{\bf 400}, 354 (1999).

\item
T.Dobzhansky,{\sl Genetics and the Origin of Species}
(Columbia Univ. Press., New York, ed.2, 1951).

\item
C.Furusawa and K.Kaneko, Emergence of Rules in Cell Society: Differentiation, Hierarchy, and 
Stability, {\sl Bull. Math. Biol.}{\bf 60}, 659(1998).

\item
D.J.Futsuyma, {\sl Evolutionary Biology} (Sinauer Associates
Inc., Sunderland, Mass, ed.2, 1986).

\item
S.J.Gould and N.Eldredge, Punctuated equilibria: the tempo and mode of evolution  reconsidered, 
{\sl Paleobiology} {\bf 3}, 115 (1977).
        
\item
L.B.Holmes, Penetrance and expressivity of limb malformations,
{\sl Birth} {Defects.} {Orig.} {Artic.} {Ser.} {\bf 15}, 321(1979).

\item
D.J. Howard and S.H.Berlocher,Ed.,
{\sl Endless Form: Species and Speciation} (Oxford Univ. Press.,
1998).

\item
K.Kaneko, Clustering, Coding, Switching, Hierarchical Ordering,
and Control in Network of Chaotic Elements,
{\sl Physica}{\bf 41D},137(1990).

\item
K.Kaneko, Relevance of Clustering to Biological Networks,
{\sl Physica}{\bf 75D},55(1994).

\item
K.Kaneko and T.Yomo, Isologous Diversification: A Theory of Cell Differentiation,
{\sl Bull.Math.Biol.}{\bf 59},139(1997).

\item
K.Kaneko and T.Yomo, Isologous Diversification for Robust Development of
Cell Society,
{\sl J. Theor. Biol.}{\bf 199},243(1999).

\item
A.Kashiwagi, T.Kanaya, T.Yomo, I. Urabe,
How small can the difference among competitors be for coexistence to occur, 
{\sl Researches on Population Ecology} {\bf 40}, in press.

\item
E.Ko, T.Yomo, I.Urabe, Dynamic Clustering of bacterial population,
{\sl Physica} {\bf 75D}, 81 (1994)

\item
C. Kobayashi et al., 
Thermal conversion from low- to high-activity forms of catalase I from 
Bacillus stearothermophilus
{\sl J. Biol. Chem.} {\bf 272}, 23011 (1997).

\item
A.S.Kondrashov and A.F.Kondrashov, Interactions among quantitative traits
in the course of sympatric speciation, 
{\sl Nature} {\bf 400},351(1999).

\item
R.Lande, Models of speciation by sexual selection on phylogenic traits,
{\sl Proc. Natl. Acad. Sci. USA} {\bf 78}, 3721 (1981).

\item
O.E. Landman, The inheritance of acquired characteristics,
Annu. Rev. Genet. {\bf 25}, 1 (1991) 

\item
J.Maynard-Smith, Sympatric Speciation, 
{\sl The American Naturalist} {\bf 100},637(1966).

\item
J.Maynard-Smith and E.Szathmary,
{\sl The Major Transitions in Evolution} (W.H.Freeman,
1995).

\item
A. Novick and M. Weiner, Enzyme induction as an all-or-none phenomenon,
Biochemistry {\bf 43}, 553 (1957).

\item
J.M.Opitz,  Some comments on penetrance and related subjects,
{\sl Am-J-Med-Genet.} {\bf 8}, 265 (1981).

\item
D.Tilman, Ecological competition between algae: Experimental confirmation of
resource-based competition theory, 
{\sl Science} {\bf 192}, 463 (1976).

\item
D.Tilman, Test of resource competition theory using four species of lake Michigan algae,
{\sl Ecology} {\bf 62}, 802(1981).

\item
G.F.Turner and M.T.Burrows, A model for sympatric speciation by sexual selection,
{\sl Proc. R. Soc. London} {\bf B 260}, 287 (1995).

\item
W.-Z.Xu, A.Kashiwagi, T.Yomo, I.Urabe, Fate of a mutant
emerging at the initial stage of evolution,
{\sl Researches on Population Ecology} {\bf 38}, 231(1996).

\end{itemize}

\pagebreak
\begin{figure}
\noindent
\epsfig{file=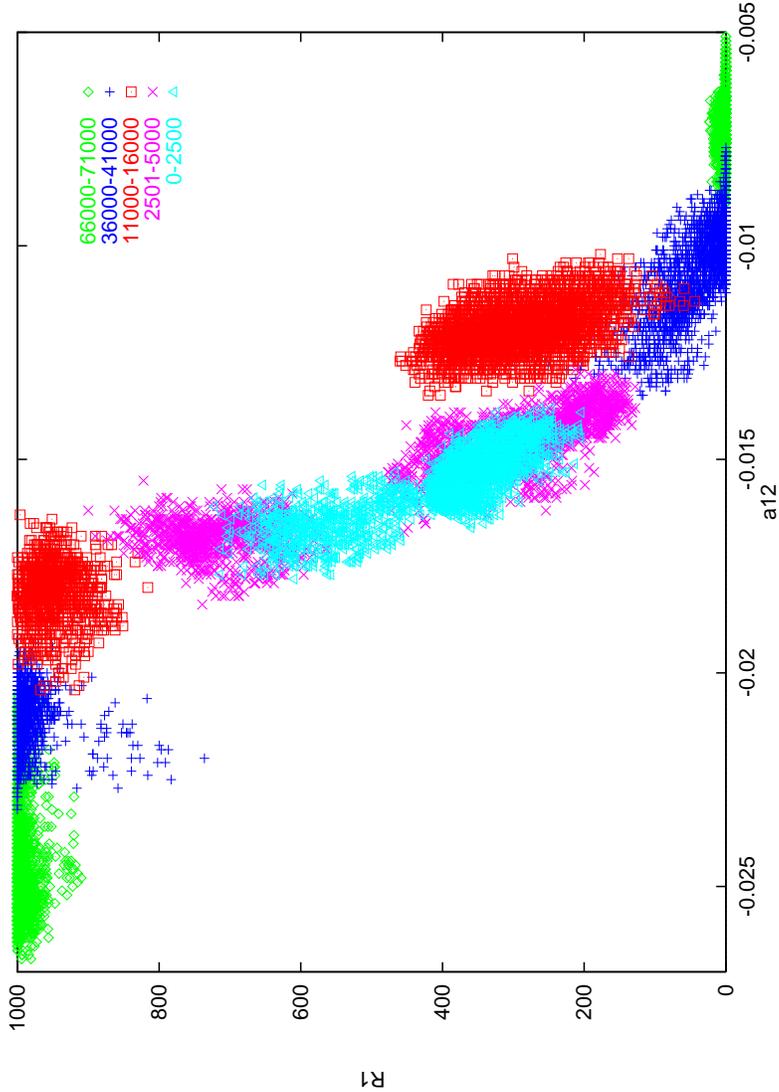,width=.75\textwidth}
\caption{
Evolution of genotype-phenotype relationship. 
In the present model, due to the nonlinear nature of the dynamics,
$x_n^{\ell}$ often oscillates in time chaotically or periodically.
Hence it is natural to use the integer part 
$R^{\ell}(j)$, as a representation of the phenotype, since it represents
the number of cyclic process used for reproduction.
Here  $(R^1,a^{12})$ is plotted for every division of individuals. 
The first 2500
divisions are plotted in light blue, divisions 2501-5000 in 
pink, 11000-16000 in red, 36000-41000 in blue, and 66000-71000 in green.
Initially, phenotypes are separated, even though the genotypes are
identical (or only slightly differ),  as shown in light blue. Later, the
genotypes are also separated, according to the difference in phenotypes.
In the simulation, the population size fluctuates around 300, after an
initial transient.  (Hence the generation number is given roughly 
by dividing this division number by 300.)
In the figures of the model, we adopt the following
parameter values and initial conditions.
The threshold number $Thr$ for the reproduction is
1000, and the mutation rate of the parameters $\epsilon$ is 0.001. 
Initially, the genotype parameters are set as $a^{ij}=-0.1/(2\pi)$.
The parameter values for Fig.1 and 2 are set at
$p_k=1.8/(2\pi)$, $s^1=8$,$s^2=7$,$s^3=2$.}
\end{figure}

\pagebreak
\begin{figure}
\noindent
\epsfig{file=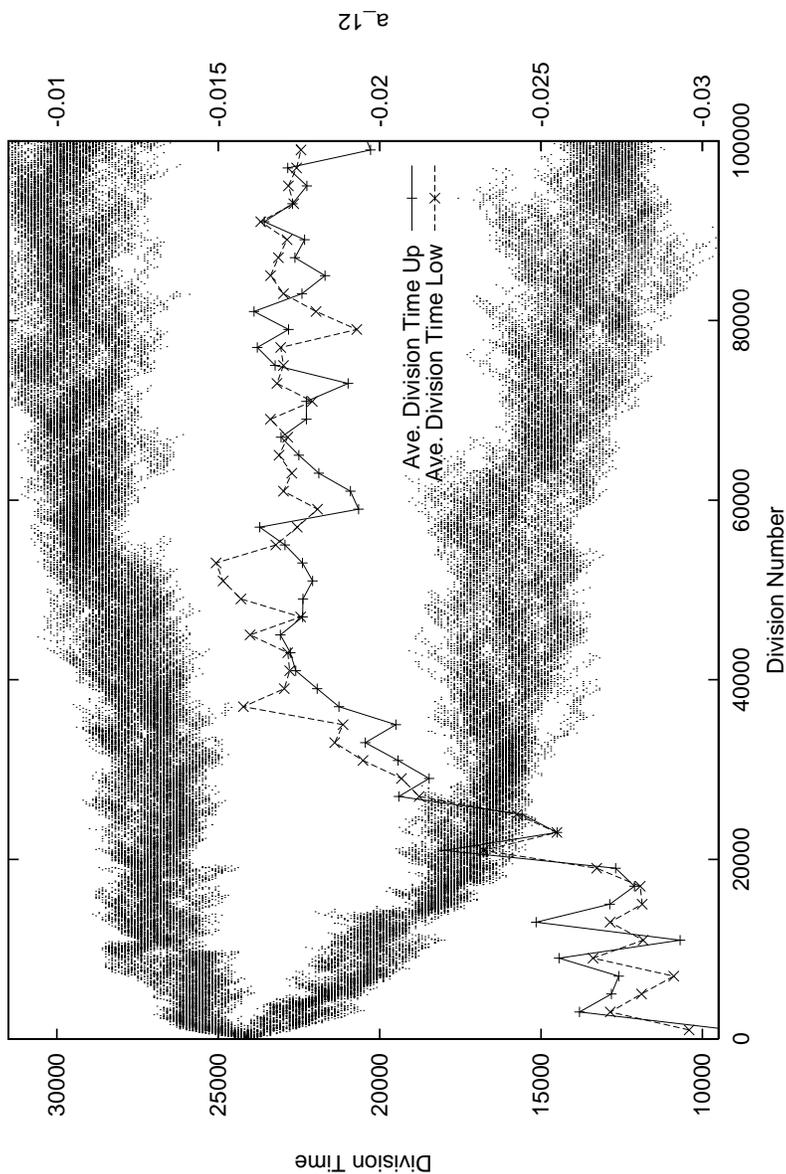,width=.8\textwidth}
\caption{
The evolution of the genotypic parameter.  The parameter $a^{12}(i)$ is
plotted as a dot at every division (reproduction) event in (a),
with the abscissa as the division number.
The average time necessary for division (reproduction) is plotted 
for the upper and lower groups, where the average 
is taken over 2000 division events (6th - 8th generation).  
As the two groups are formed around the 2000th division event, the population size
becomes twice the initial, and each division time is also
approximately doubled.  Note that the two average
division speeds of the two groups remain of the same order, even when
the genetic parameter evolves in time. }
\end{figure}

\pagebreak
\begin{figure}
\noindent
\epsfig{file=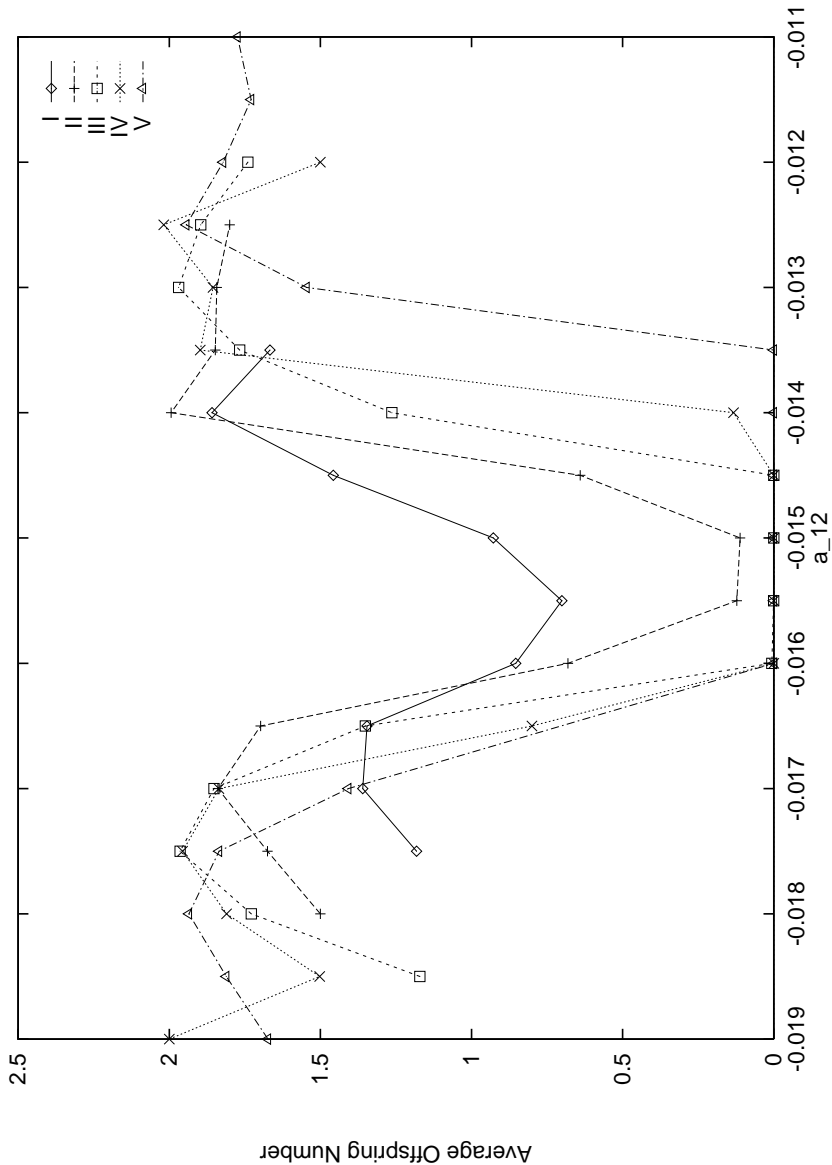,width=.8\textwidth}
\caption{
The average offspring number before death is plotted as a function of
the parameter (genotype), for simulations with sexual recombination. 
As an extension to include sexual recombination, we have also studied
a model in which two organisms satisfying the above threshold condition 
mate to reproduce two offspring.
When they mate, the offspring have parameter values that are
randomly weighted average of those of the parents, as given in the
text. 
We have measured the number of
offspring for each individual during its lifespan.   
By taking a bin width 0.005 for
the genotype parameter $a^{12}$, the average offspring number over a
given time span is measured to give a histogram.  The histogram 
over the first 7500 divisions (about 20 generations) is plotted by the
solid line (I), 
and the histogram for later divisions is overlaid with a different
line, as given by II (over 7500-15000 divisions), III
(1.5-2.25 $\times 10^4$), IV(2.25-3 $\times 10^4$), and V(3.75-4.5 $\times 10^4$). 
As shown, a hybrid offspring 
will be sterile after some generations.  
Here we have used the same model and the initial condition as in Fig.1
and imposed recombination, with the parameters
$p_k=1.5/(2\pi)$ and $s^1=s^2=s^3=2$.
In the run, the population fluctuates around 340.}
\end{figure}

\end{document}